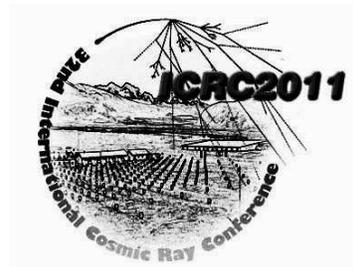

# Searching for Fast Optical Transients using VERITAS Cherenkov Telescopes

SEAN GRIFFIN, DAVID HANNA, AND ADAM GILBERT

*Physics Department, McGill University, 3600 University Street, Montreal, QC H3A 2T8, Canada*

griffins@physics.mcgill.ca

**Abstract:** Astronomical transients are intrinsically interesting things to study. Fast optical transients (microsecond timescale) are a largely unexplored field of optical astronomy mainly due to the fact that large optical telescopes are oversubscribed. Furthermore, most optical observations use instruments with integration times on the order of seconds and are thus unable to resolve fast transients. Current-generation atmospheric Cherenkov gamma-ray telescopes, however, have huge collecting areas (e.g., VERITAS, which consists of four 12-m telescopes), larger than any existing optical telescopes, and time is typically available for such studies without interfering with gamma-ray observations. The following outlines the benefits of using a Cherenkov telescope to detect optical transients and the implementation of the VERITAS Transient Detector (VTD), a dedicated multi-channel photometer based on field-programmable gate arrays. Data are presented demonstrating the ability of the VTD to detect transient events such as a star passing through its field of view and the optical light curve of a pulsar.

**Keywords:** VERITAS, hardware, high time resolution photometry

## 1 Introduction

In optical astronomy, fine time resolution observations (microsecond-timescale) are rare due to the fact that most optical detectors (such as CCDs) have long integration times, on the scale of seconds or longer. Coupled with the fact that large optical telescopes are typically oversubscribed few attempts have been made to search for fast optical transients.

Atmospheric Cherenkov detectors such as VERITAS, H.E.S.S. and MAGIC, have larger effective collection areas than the largest optical telescopes even though their optical point-spread functions are inferior. The nature of gamma-ray observations makes observing under moonlight difficult or impossible. Thus, time is typically available for other studies.

The H.E.S.S. group carried out a search for optical transients from targets such as X-ray binary systems where timescales on the order of 10-100 $\mu$s are expected. They used a purpose-built camera installed specifically for this search[1].

We have developed a rate-meter which can be integrated into the standard VERITAS telescope readout systems to allow for parasitic running and/or easy transition to optical searching duing periods when VERITAS is not being used for gamma-ray astronomy. This will allow for more extended and therefore more sensitive searches. In the following we describe this device, its implementation, and results from initial tests.

## 2 The VERITAS Transient Detector

### 2.1 The Camera

The VERITAS camera[2], shown in Fig. 1, is composed of 499 close-packed pixels, each comprising a 29-mm photo-multiplier tube and a light cone which increases the effective area and blocks off-angle light. The VERITAS Transient Detector (VTD) uses the centre seven pixels: a central pixel surrounded by a guard ring of six others. The centre pixel is used to collect light from the target and the guard ring provides the ability to veto non-astronomical transient events.

Each pixel has a field of view of $0.15°$ and the telescope's optical point spread function is $0.12°$ in diameter[3]. Thus, all light from a point source (*i.e.* a compact astronomical target) is collected within a single pixel. Non-astronomical transients, such as meteors, aircraft, or orbiting objects will appear in more than one pixel simultaneously, or will move from pixel to pixel over time. Hence, any events not fully contained within the centre pixel can be easily rejected in the offline analysis.

### 2.2 The Rate Meter

The VTD rate meter is built from a commercially available FPGA prototyping board, the Xilinx ML402 Evaluation Platform, which contains a Virtex-4 XC4VSX35-FF668-1 FPGA. It also includes a custom-built mezzanine board for loading NIM signals from the constant-fraction discriminators (CFDs) of the standard VERITAS readout into the FPGA. The electronics on the mezzanine board translate the signals into the low-voltage differential signaling (LVDS) standard. A photo of the device is shown in Fig. 2.

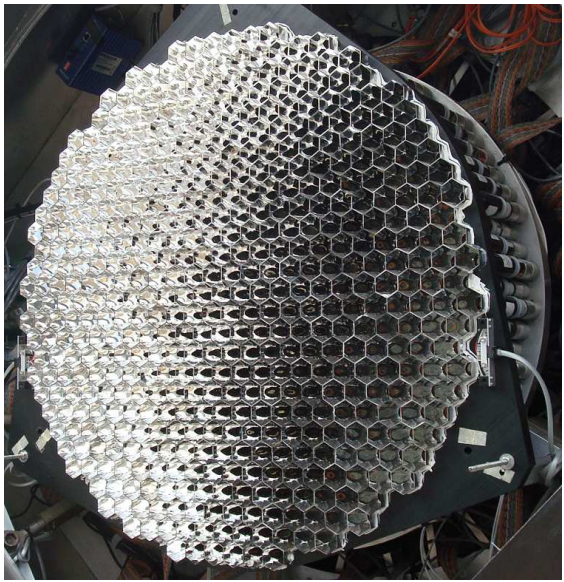

Figure 1: A VERITAS camera comprising 499 close-packed pixels. For the purposes of searching for optical transients, only the centre seven will be used.

PMT pulses are amplified and pass through 50m of cable before being fed into the CFDs. They are are triggered at a level of typically 5-7 photoelectrons and counted by firmware-defined 16-bit scalers in the FPGA which are read by the embedded microprocessor every 5.5 $\mu$s and loaded into a buffer.

The maximum count rate seen by the VTD is a function of several factors; the most important are listed here. The mezzanine electronics have a maximum trigger rate of 830 MHz, but the scalers on the FPGA are physically limited to 400 MHz. The VTD counts triggers from the CFDs, which can handle rates of about 35 MHz, which is where the bottleneck is; we are currently investigating methods of circumventing this limitation.

Once the buffer is full (70 samples from the scalers) it is sent via Ethernet to a host computer, a process which takes 61.87 $\mu$s. This transfer introduces a deadtime of about 15% although this is not deadtime in the traditional sense: the firmware scalers are still counting during buffer flushes, but the microprocessor is busy and unable to sample them. Thus, the first sample after a buffer flush contains all the counts that were taken in the interim. A block diagram of this setup is shown in Fig. 3.

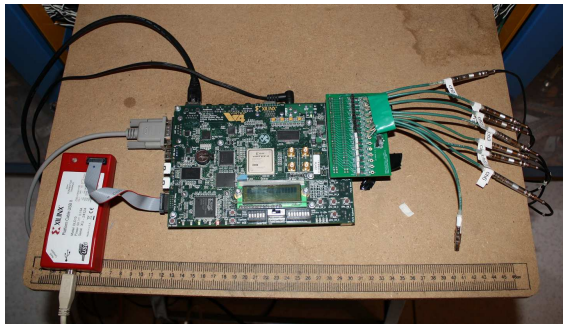

Figure 2: The VTD rate meter. The main electronics board is a Xilinx ML402 Evaluation Platform. The LEMO cables at the right connect to the standard VERITAS readout chain.

## 3 Use on VERITAS

We intend to use the VTD during periods of bright moonlight when gamma-ray observations are impractical. This allows us to explore a different part of astrophysics without compromising the primary science goals of VERITAS. The conversion to transient searches requires no more than swapping seven cables in the readout of each telescope.

## 4 Test Results

The VTD electronics have been tested using a pulse generator as well as an LED-illuminated PMT in a dark box. Following these checks, we have field-tested the device on Mt Hopkins using the Whipple 10-m telescope[4] and one of the VERITAS telescopes.

### 4.1 Drift Scans

A drift scan is a run where a star is allowed to pass through the field of view of the camera. This should be readily apparent in the data as a time-dependent signal seen in the pixel trigger rates. A scan of this type was made using the Whipple 10-m telescope and the results are shown in Fig. 4 where the count rate as a function of time is plotted for three pixels. The rise and fall of the rates in adjacent pixels indicate the passage of the target star. MHz rates are seen and the Poisson statistics associated with the corresponding samples give rise to scatter in the data. The time bins correspond to 70 samples (*i.e.* one buffer) or about 0.45 ms.

### 4.2 Crab Pulsar Detection

We have tested the VTD on a VERITAS telescope while tracking the Crab pulsar. The Crab has been seen to pulse all across the electromagnetic spectrum and observations of it pulsing at optical wavelengths are routinely used by gamma-ray experiments to verify timing and anal-ysis codes[5, 6]. In our test we recorded time stamps from the host computer along with instantaneous rates from the central pixel and used them to barycentre the photons arrival times. Using ephemerides from the Jodrell Bank radio observatory[7] we compute the phase of each rate measurement and increment the appropriate bin in a phaseogram. Due to the lack of an absolute (*i.e.* GPS) timestamp, the result is accurate to an arbitrary shift in the phase which is corrected manually afterwards.

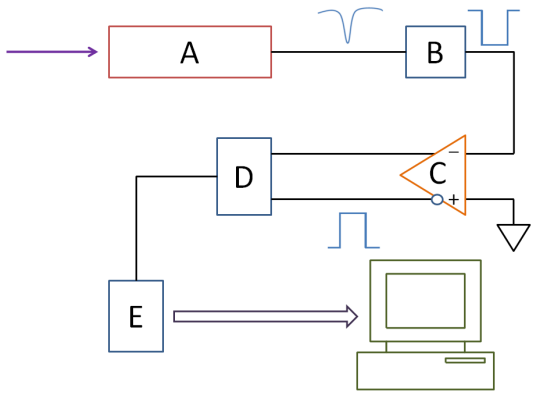

Figure 3: Block diagram of how the VTD is implemented. Photons enter a PMT (A). The signal is sent to a constant-fraction discriminator (B), the output of which is fed into a high speed LVDS (low voltage differential signaling) comparator (C) which produces positive voltage logic with signal response times on the order of 1ns. The signal pulses are counted by a 16-bit scaler (D). A microprosessor embedded in the FPGA (E) reads the scaler values every 5.5 $\mu s$ and stores them in a memory buffer which is subsequently read out over Gigabit Ethernet to a host computer. The device is hardware-limited to rates $\leq 400$ MHz.

The results from 45 minutes of observations are shown in Fig. 5 where a clear detection of the optical pulsar signal is seen. The corresponding sensitivity is approximately $30\sigma/\sqrt{hour}$.

## 5  Conclusions

The design for the VTD, a multi-channel high-time-resolution photometer, has been outlined, as has its use on VERITAS. Following the success of tests done over the past few months, searches for microecond optical flares will now take place.

The primary source candidates are X-ray binary systems where the emission mechanism for fast optical transients is likely the accretion of matter onto compact objects. For objects with sizes in the 10-km range, this suggests a typical time scale of 10-100 $\mu s$. Millisecond-timescale optical variablity has already been observed in X-ray binaries[8, 9] and pulsars[10]. Fluctuations in the optical afterglow of gamma-ray bursts, with timescales of less than an hour have also been detected[11]. It may be possible to see fine structure in the light curves of such transients.

Given the nature of when gamma-ray observations can take place, there is time available for such observations without interfering with the primary science goals of VERITAS.

## 6  Acknowledgements

VERITAS is supported by grants from the US Department of Energy, the US National Science Foundation, and the Smithsonian Institution, by NSERC in Canada, by Science Foundation Ireland. and by STFC in the UK.

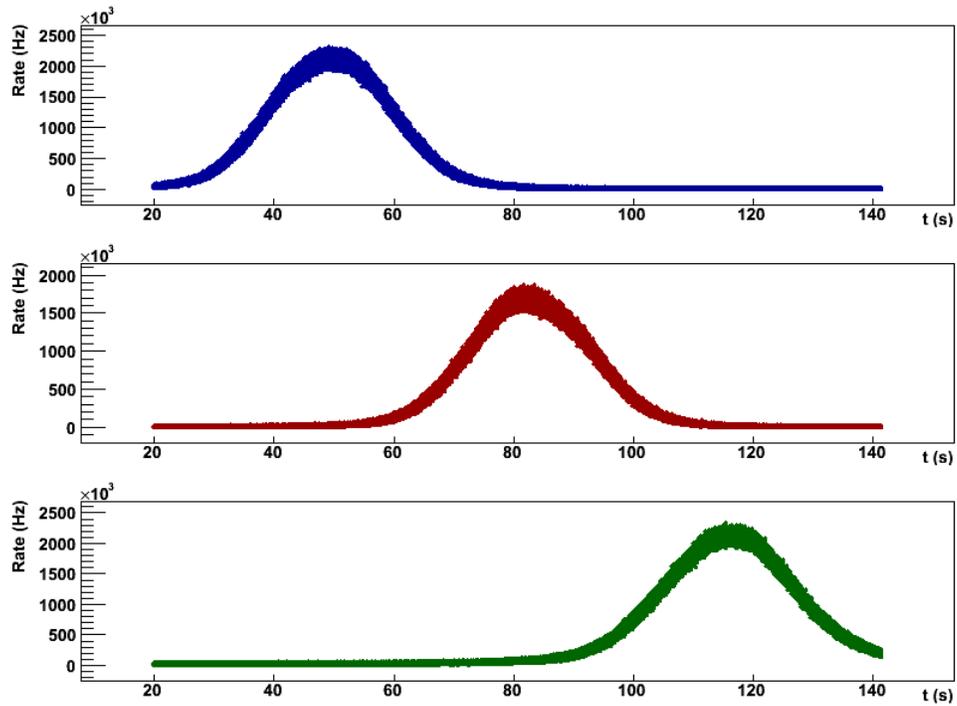

Figure 4: Result of a drift scan taken using the Whipple 10-m telescope. A star drifts across three camera pixels (order top to bottom: Channel 2 (veto ring) → Channel 0 (central pixel) → Channel 5 (veto ring)). The scatter in the data is consistent with Poissonian fluctuations.

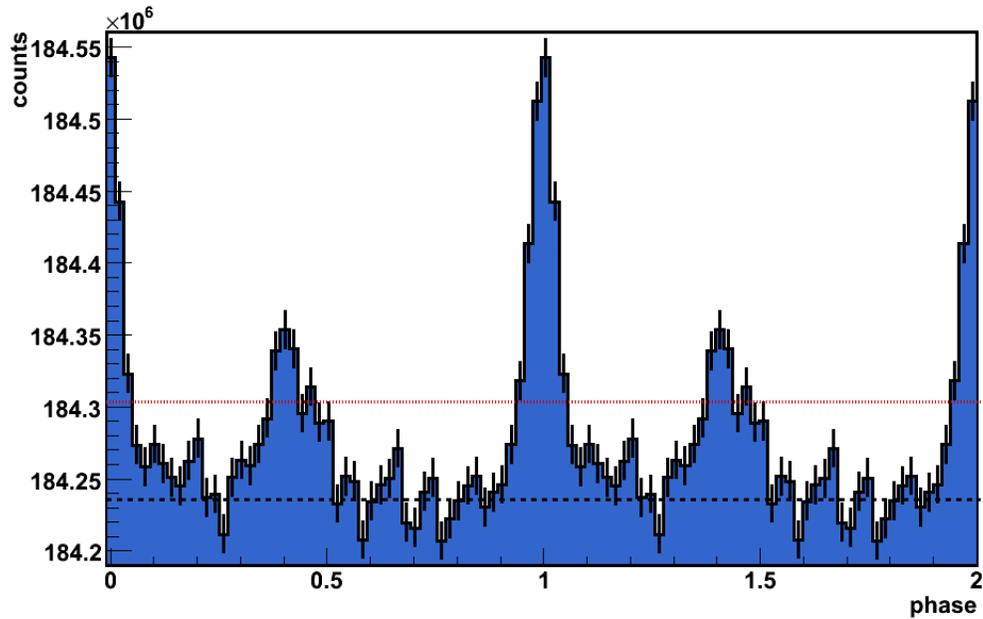

Figure 5: Phaseogram for the Crab optical pulsar. Two periods are plotted for clarity. The dashed line represents the mean rate and the dotted line represents the $5\sigma$ detection threshold. The main pulse (phase = 1) is clearly seen as is the inter-pulse (phase = 0.4).